\documentclass[a4paper,12pt, epsfig]{article}
\def\letter{0}\def\pr{0}
\pdfoutput=1 
\usepackage{epsfig}
\usepackage{epstopdf}
\usepackage{graphicx}
\usepackage{ifthen}

\usepackage{amsmath}
\usepackage[psamsfonts]{amssymb}
\usepackage{euscript}

\usepackage{latexsym}
\usepackage[arrow,matrix,curve]{xy}

\newskip\humongous \humongous=0pt plus 1000pt minus 1000pt

\newif\ifdtup

\def\,{\hspace{-.1cm}}
\def\hsp{,\hspace{.7cm}}

\def\fc#1#2 {\frac{n}{q}#1\frac{n}{q}#2}

\def\hf{H\p_{f_0}}

\newcommand{\vac}{\ensuremath{|0\rangle}}
\newcommand{\stt}{\ensuremath{|\psi\rangle}}

\renewcommand{\tanh}{\textrm{tanh}}
\newcommand{\sech}{\textrm{sech}}

\renewcommand{\theequation}{\arabic{section}.\arabic{equation}}
\renewcommand{\(}{\begin{equation}}
\renewcommand{\)}{end{equation} \vspace{-.05in}\linebreak}

\newcounter{saveeqn}
\newcounter{savealpheqn}

\newcommand{\alpheqn}{\setcounter{saveeqn}{\value{equation}}%
  \stepcounter{saveeqn}\setcounter{equation}{0}%
  \renewcommand{\theequation}{\mbox{\arabic{section}.\arabic{saveeqn}
\alph{equation}}}
  \renewcommand{\)}{\end{equation}}}
\def\part#1{\frac{\partial}{\partial{#1}}}%
\def\group#1{\refstepcounter{equation}\setcounter{saveeqn}
 {\value{equation}}%
  \label{#1}\setcounter{equation}{0}%
\renewcommand{\theequation}{\mbox{\arabic{section}.\arabic{saveeqn}
\alph{equation}}}
  \renewcommand{\)}{\end{equation}}}
\newcommand{\reseteqn}{\setcounter{equation}{\value{saveeqn}}%
  \renewcommand{\theequation}{\arabic{section}.\arabic{equation}}%
  \renewcommand{\)}{\end{equation}}}

\newcommand{\aalpheqn}{\setcounter{saveeqn}{\value{equation}}%
  \stepcounter{saveeqn}\setcounter{equation}{0}%
  \renewcommand{\theequation}{\mbox{
        \Alph{subsection}.\arabic{saveeqn}\alph{equation}}}
   \renewcommand{\)}{\end{equation}}}
\newcommand{\areseteqn}{\setcounter{equation}{\value{saveeqn}}%
  \renewcommand{\theequation}{\Alph{subsection}.\arabic{equation}}%
  \renewcommand{\)}{\end{equation}}}

\renewcommand{\thefootnote}{\alph{footnote}}
\renewcommand{\(}{\begin{equation}}
\renewcommand{\)}{\end{equation}}
\newcommand{\ba}{\begin{eqnarray}}
\newcommand{\ea}{\end{eqnarray}}

\renewcommand{\b}{\beta}

\newcommand{\cbp}{\mathop{\vtop{\ialign{##\crcr
   $\hfil\displaystyle{}\hfil$\crcr\noalign{\kern-13pt\nointerlineskip}
   \BIG{)}\hskip0pt\crcr\noalign{\kern3pt}}}}}
\newcommand{\pa}{\mathop{\vtop{\ialign{##\crcr

$\hfil\displaystyle{\oplus}\hfil$\crcr\noalign{\kern+1pt\nointerlineskip
}
   \hspace{.08in}$^{\alpha=0}$\hskip6pt\crcr\noalign{\kern3pt}}}}}
\renewcommand{\hsp}{,\hspace{.3in}}
\newcommand{\p}{^\prime}

\def\D{\ensuremath{{\cal D}}}

\catcode`\@=11
\def\vereq#1#2{\lower3pt\vbox{\baselineskip1.5pt \lineskip1.5pt
\ialign{$\m@th#1\hfill##\hfil$\crcr#2\crcr\sim\crcr}}}
\catcode`\@=12

\renewcommand{\(}{\begin{equation}}
\renewcommand{\)}{\end{equation}}

\def\pin#1{\int \frac{d#1}{2\pi}}
\def\ppin#1{\int^+ \frac{d#1}{2\pi}}
\def\pink#1{\int \frac{d^{#1}k}{(2\pi)^{#1}}}

\def\Bd#1{B^\dag_{k_{#1}}}

\def\df{\mathcal{D}_{f_0}}
\def\dF{\mathcal{D}_F}

\def\I{\mathcal{I}}

\newcommand{\beas}{\begin{eqnarray*}}
\newcommand{\eeas}{\end{eqnarray*}}

\newcommand{\bquo}{\begin{quote}}
\newcommand{\enqu}{\end{quote}}

\newcommand{\g}{{\mathfrak g}}

\def\ch{{\mathcal{H}}}

\def\ok#1{\omega_{k_{#1}}}

\def\V#1{V^{(#1)}[gf_0(x)]}

\newcommand{\beq}{\begin{equation}}
\newcommand{\eeq}{\end{equation}}
\newcommand{\bea}{\begin{eqnarray}}
\newcommand{\eea}{\end{eqnarray}}

\newskip\humongous \humongous=0pt plus 1000pt minus 1000pt

\newif\ifdtup

\catcode`\@=11

\ifthenelse{\equal{\letter}{0}}{ 

\@addtoreset{equation}{section}
\def\theequation{\arabic{section}.\arabic{equation}}

\def\@normalsize{\@setsize\normalsize{15pt}\xiipt\@xiipt
\abovedisplayskip 14pt plus3pt minus3pt%
\belowdisplayskip \abovedisplayskip
\abovedisplayshortskip \z@ plus3pt%
\belowdisplayshortskip 7pt plus3.5pt minus0pt}

\def\small{\@setsize\small{13.6pt}\xipt\@xipt
\abovedisplayskip 13pt plus3pt minus3pt%
\belowdisplayskip \abovedisplayskip
\abovedisplayshortskip \z@ plus3pt%
\belowdisplayshortskip 7pt plus3.5pt minus0pt
\def\@listi{\parsep 4.5pt plus 2pt minus 1pt
      \itemsep \parsep
      \topsep 9pt plus 3pt minus 3pt}}

\relax

\def\section{\@startsection{section}{1}{\z@}{3.5ex plus 1ex minus  .2ex}{2.3ex plus .2ex}{\large\bf}}

\def\thesection{\arabic{section}}
\def\thesubsection{\arabic{section}.\arabic{subsection}}

\def\appendix{\setcounter{section}{0}
 \def\thesection{Appendix \Alph{section}}
 \def\thesubsection{\Alph{section}.\arabic{subsection}}
 \def\theequation{\Alph{section}.\arabic{equation}}}
\renewcommand{\theequation}{\arabic{section}.\arabic{equation}}

}{
\renewcommand{\theequation}{\arabic{equation}}

} 

\begin{document}
\def\thefootnote{\fnsymbol{footnote}}
\def\thetitle{Removing Tadpoles in a Soliton Sector}
\def\autone{Jarah Evslin}
\def\auttwo{Hengyuan Guo}
\def\affa{Institute of Modern Physics, NanChangLu 509, Lanzhou 730000, China}
\def\affb{University of the Chinese Academy of Sciences, YuQuanLu 19A, Beijing 100049, China}

\title{Titolo}

\ifthenelse{\equal{\pr}{1}}{
\title{\thetitle}
\author{\autone}
\author{\auttwo}
\affiliation {\affa}
\affiliation {\affb}

}{

\begin{center}
{\large {\bf \thetitle}}

\bigskip

\bigskip


{\large \noindent  \autone{${}^{1,2}$} and \auttwo{${}^{1,2}$}}


\vskip.7cm

1) \affa\\
2) \affb\\

\end{center}

}

\begin{abstract}
\noindent
It has long been known that perturbative calculations can be performed in a soliton sector of a quantum field theory by using a soliton Hamiltonian, which is constructed from the defining Hamiltonian by shifting the field by the classical soliton solution.  It is also known that even if tadpoles are eliminated in the vacuum sector, they remain in the soliton sector.     In this note we show, in the case of quantum kinks at two loops, that the soliton sector tadpoles may be removed by adding certain quantum corrections to the classical solution used in this construction.  Stated differently, the renormalization condition that the soliton sector tadpoles vanish may be satisfied by renormalizing the soliton solution.


\end{abstract}

%
\setcounter{footnote}{0}
\renewcommand{\thefootnote}{\arabic{footnote}}

\ifthenelse{\equal{\pr}{1}}
{
\maketitle
}{}

\section{Introduction}
The eigenvalues of the Hamiltonian $H[\phi(x)]$ are the energies of the states of a theory.  If any other operator $H\p[\phi(x)]$ is related to $H[\phi(x)]$ by a similarity transformation, it will have the same eigenvalues and so it may equivalently be used to calculate energies of states.  This observation is useful because, at least in the case of quantum kinks, the energies of states in a kink sector of a quantum field theory cannot be found perturbatively using $H[\phi(x)]$ but can \cite{dhn2} be found perturbatively using the kink Hamiltonian
\beq
\hf[\phi(x)]=H[\phi(x)+f_0(x)]
\eeq
where $f_0(x)$ is the classical kink solution.  

In the case of the Sine-Gordon model, the one-loop spectrum calculated in Ref.~\cite{dhn2} was extended to two loops in Refs.~\cite{vega,verwaest}.  Here it was found that, although tadpoles were eliminated in the vacuum sector by a choice of renormalization conditions, tadpole diagrams yield important contributions to the ground state energy of the kink at two loops.  In Ref.~\cite{menormal,meshape} it was shown that the same is true of the energies of excited states.  These tadpoles do not lead to any inconsistencies.  However the tadpoles appear in most diagrams, leading one to wonder whether perturbation theory may be simplified by eliminating them, and thus eliminating most diagrams. 

In this note we will introduce a quantum kink Hamiltonian
\beq
H\p_F[\phi(x)]=H[\phi(x)+F(x)]
\eeq
where $F(x)$ is the classical kink solution plus perturbative corrections
\beq
F(x)=\sum_{n=0}g^{n} f_n(x).
\eeq
Here all $f_n(x)$, like the original $f_0(x)$, are of order $O(1/g)$.  We will see that $f_1(x)$ necessarily vanishes and will construct the $f_2(x)$ which eliminates the leading order tadpoles.  We see no obstruction to eliminating the higher order tadpoles by fixing all of the $f_n(x)$.

Are we allowed to shift the Hamiltonian by a function that is not the classical solution?  Recall that the new Hamiltonian will have the same spectrum as the old Hamiltonian if the two are similar.  As was shown in Ref.~\cite{mekink}, this similarity holds for {\it{any}} real function $F(x)$ as
\beq
H\p_F[\phi(x)]=\dF^\dag H[\phi(x)]\dF\hsp
\dF={\rm{exp}}\left(-i\int dx F(x)\pi(x)\right). \label{hp}
\eeq
The unitary displacement operator $\dF$ shifts $\phi(x)$ by $F(x)$.

Note that the energies of the states are the spectrum of the {\it{regularized}} Hamiltonian.  Therefore $H[\phi(x)]$ needs to be the regularized Hamiltonian.   If $H[\phi(x)]$ is regularized via normal ordering, then (\ref{hp}) is satisfied if and only if $H\p_F[\phi(x)]$ is also normal ordered \cite{mekink}.  We remind the reader that in 1+1 dimensional scalar field theories, normal ordering is sufficient to eliminate ultraviolet divergences.  More generally, Eq.~(\ref{hp}) can be used as a definition of $H\p$: given a regularized $H$, it fixes the regularized $H\p$.

While this theory is UV finite, there may still be IR divergences.  For example, in the Sine-Gordon theory at three loops and the $\phi^4$ theory at two loops, the vacuum has a finite energy density and so an infinite energy.  As a result the kink state also has an infinite energy, and the kink mass is the difference between these two infinite energy levels.  This IR divergence is removed in Ref.~\cite{mephi4massa} by including a constant counterterm in the Hamiltonian density which sets the vacuum energy to zero.

We begin in Sec.~\ref{revsez} with a review of perturbation theory in the kink sector.  We describe how the classical solution may be used to define a kink Hamiltonian which has the same spectrum as the original Hamiltonian, which defines the theory.  Next in Sec.~\ref{statsez} we describe how the cubic term in the kink Hamiltonian yields a tadpole after switching from plane wave normal ordering to normal mode normal ordering.  We then describe how the construction of the kink Hamiltonian may be perturbed, yielding the construction of a new operator which we call the {\it{quantum kink Hamiltonian}}.  This new Hamiltonian again has the same spectrum.  We fix the perturbation at leading order by requiring it to cancel the leading order tadpole resulting from the original kink Hamiltonian.  In the Appendix we perform a consistency check, showing that the two-loop kink mass derived using the quantum kink Hamiltonian agrees with that derived using the original kink Hamiltonian.  The notation is summarized in Table~\ref{notab}.  

\section{Review} \label{revsez}

\begin{table}
\begin{tabular}{|l|l|}
\hline
Operator&Description\\
\hline
$\phi(x),\ \pi(x)$&The real scalar field and its conjugate momentum\\
$A^\dag_p,\ A_p$&Creation and annihilation operators in plane wave basis\\
$B^\dag_k,\ B_k$&Creation and annihilation operators in normal mode basis\\
$\phi_0,\ \pi_0$&Zero mode of $\phi(x)$ and $\pi(x)$ in normal mode basis\\
$::_a,\ ::_b$&Normal ordering with respect to $A$ or $B$ operators respectively\\
$H$&The defining Hamiltonian\\
$\hf$&$\df$-transformed $H$\\
$H\p_F$&$\D_F$-transformed $H$\\
$H_n,\ H_n^F$&The $g^{n-2}$ term in $\hf$ and $H\p_F$\\
\hline
Symbol&Description\\
\hline
$f_0(x),\ f_2(x)$&The classical kink solution and its leading quantum deformation\\
$F(x)$&The deformed/quantum kink solution\\
$\df$&Unitary operator that translates $\phi(x)$ by the classical kink solution\\
$\D_F$&Unitary operator that translates $\phi(x)$ by the quantum kink solution $F(x)$\\
${\g}_B(x)$&The kink linearized translation mode\\
${\g}_k(x),\ {\g}_S(x)$&Continuum and discrete normal mode\\
$\gamma_i^{mn}$&Coefficient of $\phi_0^m B^{\dag n}\vac_0$ in order $i$ eigenstate of $\hf$\\
$V_{ijk}$&Derivative of the potential contracted with various functions\\
$\I(x)$&Contraction factor from Wick's theorem\\
$p$&Momentum\\
$k$&The analog of momentum for normal modes\\
$\omega_k,\ \omega_p$&The frequency ($\sqrt{M^2+k^2}$ or $\sqrt{M^2+p^2}$) corresponding to $k$ or $p$\\
$Q_n$&$n$-loop correction to kink ground state energy in the kink Hamiltonian $\hf$\\
\hline
State&Description\\
\hline
$\vac\ (\vac_i)$&Kink ground state as eigenvector of $\hf$ (at order $i$)\\
$\vac^K\ (\vac^K_i)$&Kink ground state as eigenvector of $H\p_K$ (at order $i$)\\
\hline

\end{tabular}
\caption{Summary of Notation}\label{notab}
\end{table}

Let us begin with a (1+1)-dimension scalar field theory defined by the Hamiltonian
\bea
H&=&\int dx \ch(x) \label{hd}\\
\ch(x)&=&\frac{1}{2}:\pi(x)\pi(x):_a+\frac{1}{2}:\partial_x\phi(x)\partial_x\phi(x):_a+\frac{1}{g^2}:V[g\phi(x)]:_a.\nonumber
\eea
The plane wave normal ordering $::_a$ will be defined momentarily.  
Consider a stationary solution $f_0(x)$ of the classical equations of motion
\beq
\phi(x,t)=f_0(x)\hsp
-gf_0(x)^{''}+\V{1}=0
\label{fd}
\eeq
where $\V{n}$ is the $n$th derivative of $V[g\phi(x)]$ with respect to its argument $g\phi(x)$ evaluated at $\phi(x)=f_0(x)$.
Using Eq.~(\ref{hp}) we may calculate the corresponding kink Hamiltonian $\hf$
\bea
\hf&=&\df^\dag H\df=Q_0+\sum_{n=2}^{\infty}H_n\hsp
H_{n(>2)}=\frac{g^{n-2}}{n!}\int dx \V{n}:\phi^n(x):_a\label{hfe}\\
H_2&=&\frac{1}{2}\int dx\left[:\pi^2(x):_a+:\left(\partial_x\phi(x)\right)^2:_a+\V{2}:\phi^2(x):_a\right.]\nonumber
\eea
where $Q_0$ is the mass of the classical kink and $M^2$ is defined to be $V^{(2)}[g f_0(\pm\infty)].$  Note that if $f_0(+\infty)\neq f_0(-\infty)$ then the quantum kink will accelerate towards the lower energy vacuum and so there is no corresponding Hamiltonian eigenstate and thus no eigenvalue to calculate \cite{weigelstab}.

Inserting the constant frequency Ansatz
\beq
\phi(x,t)=e^{-i\omega t}\g(x)
\eeq
into the linearized wave equation derived from $H_2$ one finds the Sturm-Liouville equation of motion
\beq
\V{2}{\g}(x)=\omega^2{\g}(x)+{\g}^{\prime\prime}(x). \label{sl}
\eeq
In general it has three kinds of solutions.  There will be a zero-mode ${\g}_B(x)$ with $\omega_B=0$ as a result of the translation invariance of the Hamiltonian.  There will be continuum modes ${\g}_k(x)$ with $\ok{}>M$ and $k$ defined up to a sign by
\beq
\ok{}=\sqrt{M^2+k^2}. \label{ok}
\eeq
Finally there will be discrete modes ${\g}_S(x)$ with $0<\omega_S<M$.  We will refer to all three kinds of solutions as normal modes.

Adopting the convention\footnote{We have assumed that $V[gf_0(x)]$ is symmetric about the center of the vortex, a choice which eliminates various classical \cite{tamasstab} and quantum \cite{weigelstab} instabilities.  However the generalization to an arbitrary $V[\phi]$ is straightforward.   In this generalization, one removes ${\g}_k(-x)$ from this list of equalities.}
\beq
{\g}_k(-x)={\g}_k^*(x)={\g}_{-k}(x),
\eeq
we  normalize the normal modes by imposing
\beq
\int dx |{\g}_{B}(x)|^2=1,\
\int dx {\g}_{k_1} (x) {\g}^*_{k_2}(x)=2\pi \delta(k_1-k_2),\ 
\int dx {\g}_{S_1}(x){\g}^*_{S_2}(x)=\delta_{S_1S_2}.
\eeq
Then, using a fundamental result of Sturm-Liouville theory, the normal modes satisfy the completeness relations
\beq
{\g}_B(x){\g}_B(y)+\ppin{k}{\g}_k(x){\g}^*_{k}(y)=\delta(x-y) \label{comp}
\eeq
where we have defined $\int^+$ to be the integral over continuum modes plus the sum over discrete nonzero normal modes
\beq
\ppin{k}=\pin{k}+\sum_S.
\eeq

So far our discussion has been classical.  Let us now introduce the Schrodinger picture quantum field $\phi(x)$ and its conjugate momentum $\pi(x)$.  These are independent of time and so, as usual, one may expand the scalar field and its conjugate momentum in a plane wave basis \bea
\phi(x)&=&\pin{p}\left(A^\dag_p+\frac{A_{-p}}{2\omega_p}\right) e^{-ipx}\label{adec}\\
 \pi(x)&=&i\pin{p}\left(\omega_pA^\dag_p-\frac{A_{-p}}{2}\right) e^{-ipx}.
\nonumber
\eea
However the completeness of the normal modes means that any field may be expanded in the normal mode basis \cite{cahill76}
\bea
\phi(x)&=&\phi_0 {\g}_B(x) +\ppin{k}\left(B_k^\dag+\frac{B_{-k}}{2\omega_k}\right) {\g}_k(x)\label{bdec}\\
\pi(x)&=&\pi_0 {\g}_B(x)+i\ppin{k}\left(\omega_kB_k^\dag - \frac{B_{-k}}{2}\right) {\g}_k(x).\nonumber
\eea
The two bases are related by a Bogoliubov transformation.

The canonical algebra $[\phi(x),\pi(y)]=i\delta(x-y)$ together with the completeness relations  of the plane wave and normal mode bases can then be used to fix the commutators of these component operators
\bea
[A_p,A_q^\dag]&=&2\pi\delta(p-q)\\
{[\phi_0,\pi_0]}&=&i\hsp
[B_{S_1},B^\dag_{S_2}]=\delta_{S_1S_2}\hsp
[B_{k_1},B^\dag_{k_2}]=2\pi\delta(k_1-k_2).\nonumber
\eea
Note that $A_p^\dag$ is the adjoint of $A_{p}/(2\omega_p)$, and $B_k^\dag$ is the adjoint of $B_{k}/(2\ok{})$.

Any Schrodinger picture operator constructed from $\phi(x)$ and $\pi(x)$ may then be decomposed in terms of either the plane wave basis or the normal mode basis.   There is a natural definition of normal ordering corresponding to each decomposition.  We will use $::_a$ to denote the plane wave normal ordering defined by using the plane wave decomposition (\ref{adec}) of all operators and then placing all $A^\dag$ on the left.  The normal mode normal ordering $::_b$ is defined by first using the normal mode decomposition (\ref{bdec}) and then placing all $B^\dag$ and $\phi_0$ on the left.



In Ref.~\cite{cahill76}, it was noted that just as the plane wave decomposition (\ref{adec}) diagonalizes the free theory describing the linearized vacuum sector, the normal mode decomposition diagonalizes the linearized kink Hamiltonian $H_2$
\bea
H_2&=&Q_1+\frac{\pi_0^2}{2}+\ppin{k}\omega_k B^\dag_k B_k \label{h2a}\\
Q_1&=&-\frac{1}{4}\ppin{k}\pin{p}\frac{(\omega_p-\omega_k)^2}{\omega_p}\tilde{{\g}}^2_{k}(p)-\frac{1}{4}\pin{p}\omega_p\tilde{{\g}}_{B}(p)\tilde{{\g}}_{B}(p)\nonumber
\eea
where we have defined the inverse Fourier transform
\beq
\tilde{{\g}}(p)=\int dx {\g}(x) e^{ipx}.
\eeq
Note that, in the case of continuum modes, $\tilde{{\g}}^2_{k}(p)$ contains a $\delta^2(p-k)$.   In Eq.~(\ref{h2a}) this term is multiplied by a double zero in $p-k$ and so it vanishes by the usual limit argument.   However, without recourse to ill-defined squares of delta functions, it has been shown directly \cite{memassa} that such contributions to $Q_1$ vanish at an earlier step in the derivation of Eq.~(\ref{h2a}).

As $H_2$ is a sum of quantum harmonic oscillators, the exact one-loop spectrum is now clear.  The one-loop\footnote{Although we have not used any Feynman diagrams, this correction is at one loop because $Q_1/Q_0$ is of order $O(g^2\hbar)$.} quantum correction to the ground state energy is $Q_1$.  One can also easily read the corresponding states off of (\ref{h2a}).  The one-loop ground state $\vac_0$ is the state annihilated by $H_2-Q_1$ and so by $\pi_0$ and all $B_k$ and $B_S$.
\beq
\pi_0\vac_0=B_k\vac_0=B_S\vac_0=0. \label{v0}
\eeq
Excited states $\stt_0$, at one-loop, are easily created by exciting normal modes with $B^\dag$ and boosting with $e^{i\phi_0 k/\sqrt{Q_0}}$.


The one-loop state $\stt_0$ serves as the starting point of our perturbation theory in $g$.  Using the fact that $Q_0$ is of order $O(1/g^2)$ we expand a state as
\beq
\stt=\sum_{i=0}^\infty \stt_{i},\
\stt_i=\sum_{m,n=0}^\infty \stt_{i}^{mn},\
\stt_i^{mn}=Q_0^{-i/2}\pink{n}\gamma_i^{mn}(k_1\cdots k_n)\phi_0^m\Bd1\cdots\Bd n\vac_0 \label{gesp}
\eeq
where the $n$-loop state includes all $i$ up to $i=2n-2$.  At each order $j$, the $\hf$ eigenvalue equation is
\beq
\sum_{i=0}^j \left(H_{j+2-i}-E_{\frac{j-i}{2}+1}\right)\stt_i=0. \label{ph}
\eeq
The order $j=0$ equation was solved above.  In the rest of this note we will focus on the $j=1$ equation
\beq
0=H_3\stt_0+(H_2-E_1)\stt_1=H_3\stt_0+\left(\frac{\pi_0^2}{2}+\ppin{k}\omega_k B^\dag_k B_k\right)\stt_1. \label{h3}
\eeq

\section{Canceling the Tadpole}  \label{statsez}

\subsection{The Leading Order Tadpole}
The eigenvalue equation (\ref{ph}) resembles the familiar expression from vacuum sector perturbation theory, except that the states are built from a finite number of normal mode creation operators $B^\dag$ on the one-loop kink ground state $\vac_0$ which is annihilated not by $A_p$ but instead by $B_k$ and $B_S$.  As a result the kink sector perturbation theory resembles the familiar vacuum sector perturbation theory, with the plane wave creation and annihilation operators and their normal ordering replaced by the corresponding normal mode creation and annihilation operators with their normal ordering.  Therefore, as shown in Ref.~\cite{menormal}, tadpoles arise when the normal mode normal ordered $H\p$ contains a term linear in $\phi(x)$.  

As is usual in perturbation theory, there will be more complicated tadpoles at higher orders.  These may be calculated systematically using the formalism described in Ref.~\cite{metwoloop} and we believe that the calculation below may be extended to cancel these higher order tadpoles as well.  However in this note we will restrict our attention to the leading order tadpoles, which appear explicitly in the normal mode normal ordered $H\p$.



At one loop, the $H\p$ eigenstates $\stt_0$ were fixed by $H_2$.  At the next order, one sees from (\ref{h3}) that $\stt_1$ also depends on $H_3$.   $H_3$ is given in (\ref{hfe}), however the expression there is plane wave normal ordered.  We have argued that computations are simpler if one first normal mode normal orders.  Plane wave normal ordering can be converted to normal mode normal ordering using the Wick's theorem of Ref.~\cite{mewick}
\beq
:\phi^n(x):_a=\sum_{m=0}^{\lfloor\frac{n}{2}\rfloor}\frac{n!}{2^m m!(n-2m)!}\I^m(x):\phi^{n-2m}(x):_b \label{wick}
\eeq
where $\I(x)$ is 
\beq
\I(x)=\pin{k}\frac{\left|{\g}_{k}(x)\right|^2-1}{2\omega_k}+\sum_S \frac{\left|{\g}_{S}(x)\right|^2}{2\omega_k}.\label{di}
\eeq
Note that the $k$ integral converges as $\left|{\g}_{k}(x)\right|^2$ tends to $1$ at large $|x|$.  For example, in the $\phi^4$ theory described by the potential
\beq
\frac{\phi^2(x)}{4}\left(g\phi(x)-\b\sqrt{8}\right)^2
\eeq
with the kink solution
\beq
f_0(x)= \frac{\b\sqrt{2}}{g}\left(1+\tanh(\b x)\right). \label{f}
\eeq
one finds \cite{mephi4massa}
\beq
\I(x)=\frac{1}{4\sqrt{3}}\sech^2(\b x)\tanh^2(\b x)-\frac{3}{8\pi}\sech^4(\b x). 
\eeq
Inserting (\ref{wick}) into (\ref{hfe}) one finds
\beq
H_3=g\int dx\left[\frac{\V{3}}{6}:\phi^3(x):_b+\frac{\V{3}}{2}\I(x):\phi(x):_b\right]. \label{h3b}
\eeq
The linear term on the right implies that there will indeed be tadpoles in the kink sector, as has long been appreciated \cite{vega}.

\subsection{The Quantum Kink Hamiltonian}

We would like to cancel the last term in (\ref{h3b}).  This could be canceled if the plane wave normal ordered form of $H\p$ had a term of the form $-g\V{3}\I(x)\phi(x)/2$.  The plane wave normal ordered form has no linear term because $f_0(x)$ satisfies the equations of motion.  The use of another classical ($O(1/g)$) profile which does not satisfy the equations of motion (\ref{fd}) would have resulted in a tadpole of order $O(1/g)$.  

To arrive at a tadpole of order $O(g)$, we will deform our solution $f_0(x)$ by a deformation $g^2f_2(x)$ such that $g^2f_2(x)/f_0(x)$ is of order $O(g^2)$.  This should be thought of as a quantum deformation, as restoring factors of $\hbar$ the dimensionless coupling is $g\sqrt{\hbar}$ and so $g^2\hbar f_2(x)/f_0(x)$ is of order $O(g^2\hbar)$.

More precisely, again setting $\hbar=1$, we will replace the classical solution $f_0(x)$ by the function
\beq
F(x)=f_0(x)+g^2f_2(x)
\eeq
where $f_2(x)$, like $f_0(x)$, is order $O(1/g)$.  Inserting this into (\ref{hp}) one finds the quantum kink Hamiltonian
\beq
H\p_F[\phi(x)]=H[\phi(x) + f_0(x) + g^2 f_2(x)]=\sum_{i=0}^\infty H^F_i
\eeq
where each $H^F_i$ has a coefficient of order $O(g^{i-2})$ when written in terms of $\phi(x)$ and $gf_n(x)$.  Recall that the leading correction beyond one loop involves only $i\leq 3$.  As the $f_2$ term is suppressed by two powers of $g$, the leading two terms in $H\p_F$ are identical to those in $\hf$
\beq
H^F_0=H_0=Q_0\hsp H^F_1=H_1=0.
\eeq
At order $O(g^0)$ one finds a contribution from $H[F(x)]$ of
\beq
H^F_2-H_2=\int dx gf_2(x) \left[-gf_0^{\prime\prime}(x)+\V{1}\right]=0
\eeq
where the term in brackets vanishes as a result of the classical equation of motion (\ref{fd}).  Had we included $f_1(x)$ in the choice of $F(x)$, a nontrivial contribution here would have complicated the one-loop problem.

The lowest order nonvanishing term in $H\p_F-\hf$ is therefore
\beq
H^F_3-H_3=g \int dx \phi(x) \left[-gf_2^{\prime\prime}(x)+\V{2}g f_2(x)
\right].
\eeq
The term in parentheses does not vanish, instead we may identify it as the Sturm-Liouville operator from the equation of motion for the normal modes (\ref{sl}).  This means that we may simplify the expression by expanding the function $f_2(x)$ in a basis of normal modes
\beq
f_2(x)=c_B {\g}_B(x)+\ppin{k} c_k {\g}_k(x) \label{f2exp}
\eeq
so that, using (\ref{sl}), 
\bea
H^F_3&=&H_3+g^2 \int dx \phi(x)\ppin{k} c_k \ok{}^2{\g}_k(x)\\
&=&\mathcal{T}+g\int dx \frac{\V{3}}{6}:\phi^3(x):_b
\nonumber
\eea
where the tadpole is
\bea
\mathcal{T}&=&g\int dx\left[\frac{\V{3}}{2}\I(x)+g\ppin{k} c_k \ok{}^2{\g}_k(x)\right]\phi(x)\label{tres}\\
&=&g\phi_0\int dx\left[\frac{\V{3}}{2}\I(x){\g}_B(x)\right]\nonumber\\
&&+g\ppin{k}\left[gc_{-k}\ok{}^2 +\int dx\frac{\V{3}}{2}\I(x){\g}_k(x)
\right]\left(B_{k}^\dag+\frac{B_{-k}}{2\omega_k}\right).
\nonumber
\eea

As we have already restricted our attention to symmetric kinks, a class which includes Sine-Gordon and $\phi^4$ kinks, $\V{3}$ is antisymmetric while $\I(x)$ and $\g_B(x)$ are symmetric and so the $\phi_0$ term vanishes.  Therefore, at this order, the choice of $c_B$ is irrelevant.  It simply shifts the midpoint of the kink.

The nonzero mode part of the tadpole vanishes if one fixes
\beq
c_k=-\frac{1}{2g\ok{}^2}\int dx\V{3}\I(x){\g}_{-k}(x)=-\frac{V_{\I,-k}}{2g^2\ok{}^2}
\eeq
where we have defined the notation
\beq
V_{\I k}=\int dxg\V{3}\I(x){\g}_{k}(x).
\eeq
Substituting $c_k$ into (\ref{f2exp}) one arrives at
\beq
f_2(y)=-\int dx\V{3}\I(x)\ppin{k}\frac{{\g}_{-k}(x)\g_k(y)}{2g\ok{}^2}. \label{main}
\eeq
This is our main result.  It is a formula for the quantum correction $g^2f_2$ to the classical solution $f_0$ of a symmetric kink such that the Hamiltonian $H\p_F$ obtained by shifting the field $\phi$ in the defining Hamiltonian $H$ by the quantum corrected solution $F=f_0+g^2f_2$ has no linear term when normal mode normal ordered.  In other words, it is the correction to the classical solution which cancels the leading order kink sector tadpole in (\ref{h3b}).


\section{Remarks}
The kink Hamiltonian $\hf$ describes fluctuations about the classical kink solution $f_0(x)$.   It can be used to perform perturbative calculations in the kink sector.   Even if the theory is regularized and renormalized so as to remove tadpoles in the vacuum sector, tadpoles appear in the kink sector \cite{vega}.  

In this draft we have introduced the quantum kink Hamiltonian which describes fluctuations about a quantum-corrected kink solution $F(x)=f_0(x)+g^2\hbar f_2(x)$.  We found that with the choice (\ref{main}) for $f_2(x)$, leading order tadpoles are removed.  At each order one has a function of degrees of freedom that can be added to $F(x)$ and so in principle one may impose an additional tadpole cancellation.  For example one may impose that a given one-point function vanishes order by order.  

Of course this naive counting easily allows a failure of tadpole cancellation in some finite number of quantities.  For example here the cancellation of the tadpole in the zero-mode seemed to require a symmetric potential.  This in fact is physically reasonable, one may expect an asymmetric potential to slightly shift the kink position, a shift which could manifest itself as a tadpole in the zero mode.

A generalization to asymmetric kinks would be desirable, as it would allow applications to problems of current interest, such as spectral walls \cite{muri1,muri2}.

The big question is:  Just what is the physical significance of this quantum corrected kink solution?  One may view the no tadpole condition in each topological sector as a renormalization condition, and so interpret the corresponding $F(x)$ in each sector as a renormalized soliton solution.  But does the function $F(x)$ correspond to some physically relevant observable?  In the future we will compare it to the Fourier transform of the form factors, which are known \cite{weiszff,karowskiff,babujianff} for the Sine-Gordon model, and try to make contact with the recent breakthrough \cite{andyff1,andyff2} in high momentum form factors in more general models.

Why might $F(x)$ be related to a form factor?  Recall that the kink ground state $\vac$ is an eigenstate of the kink Hamiltonian $H\p_F$, corresponding to the eigenstate
\beq
|K\rangle=\D_F^\dag\vac
\eeq
of the defining Hamiltonian $H$.   Similarly the leading order kink ground state $\vac_0$ can be used to define the state
\beq
|K\rangle_0=\D_F^\dag\vac.
\eeq
Note, in this definition we have not truncated $F(x)$ to its leading order contribution $f_0(x)$, nor have we constrained the higher order contributions.   Now a simple computation shows
\beq
{}_0\langle K|\phi(x)|K\rangle_0={}_0\langle 0|\D_F \phi(x)\D_F^\dag\vac_0={}_0\langle 0|\phi(x)+F(x)\vac_0=F(x){}_0\langle 0\vac_0+\g_B(x)\langle 0|\phi_0\vac_0.
\eeq
This expression is infinite and rather ill-defined.  

The basic problem, described long ago in Ref.~\cite{gj75}, is that unlike the true vacuum, the state $\vac_0$ is part of a continuum of states labeled by the kink momentum.  Therefore it is not normalizable, it is at best $\delta$-function normalizable and so its norm is infinite.  Similarly the term $\langle 0|\phi_0\vac_0$ is likely divergent.  As described there, this problem may be avoided by replacing the state $\vac$ with a localized wave packet $|x_0\rangle$ chosen such that
\beq
\langle x_0|x_0\rangle=1\hsp \langle x_0|\phi_0|x_0\rangle=0.
\eeq
Then, defining the state 
\beq
|K\rangle_{x_0}=\D_F^\dag|x_0\rangle
\eeq
one arrives at the intuitive formula
\beq
{}_{x_0}\langle K|\phi(x)|K\rangle_{x_0}=F(x)+\ppin{k} \g_k(x) \langle x_0|\left(B_k^\dag+\frac{B_{-k}}{2\omega_k}\right) |x_0\rangle. \label{intf}
\eeq
Thus the arbitrary function $F(x)$ is identified with the kink profile with a correction given by the matrix element in the last term.

It may seem to be a disaster that quantum kink profile $F(x)$ is arbitrary, but let us press on.  What is this matrix element?  If we expand our state $|x_0\rangle$ as in Eq.~(\ref{gesp}), one can see that at leading order the matrix element is just $\gamma_1^{01}$ times the infinite norm of $\vac$.   In the Appendix we will see that the tadpole-canceling $f_2(x)$ in (\ref{main}) is distinguished as the function which makes the dynamical part of $\gamma_1^{01}$ in the kink ground state vanish, leaving only an irreducible part which is mandated by translation-invariance.  This is similar to the familiar story in the vacuum sector of quantum field theory, where tadpole cancellation is the vanishing of a one-point Green's function which has the same form as this matrix element.  Of course we are not in the kink ground state, we are considering the wave packet state $|x_0\rangle$, so the matrix element will contain additional contributions from the excitations.

Therefore we conclude that $F(x)$ is only a reasonable approximation to the quantum kink profile ${}_{x_0}\langle K|\phi(x)|K\rangle_{x_0}$ when two conditions are satisfied.  First of all, it must be chosen as in (\ref{main}) to cancel the tadpole.  Second, the wave packet state $|x_0\rangle$ must not have the various oscillator modes too excited, or else the matrix elements (\ref{intf}) will again be large.   This second requirement is physically reasonable, exciting the oscillator modes too strongly should change the profile of the kink.

We have thus sketched an argument that $F(x)$ should be related to a quantum kink profile.  But is it then the same function already found in Refs.~\cite{gj75,gjs}?  Inserting Eq. (3.4) of \cite{gj75} into their tadpole cancellation condition Eq.~(3.12) one finds that this expression is equal to the vanishing of the bracket in the last line of our (\ref{tres}) with one difference.  Instead of the Wick contraction factor $\I(x)$, the authors find a factor which depends on their choice of perturbation $J(x)$.  This is to be expected, as our theory is plane wave normal ordered from the beginning, a factor of $\I(x)$ enters which converts plane wave normal ordering to normal mode ordering.  On the other hand, the theory used in Ref.~\cite{gj75} contains ultraviolet divergences, and physical answers arise only after these have been subtracted.  

An even more striking similarity is found between our $f_2$ and that of Eq.~(6.6) of Ref.~\cite{gjs}.  Again, this differs from our expression only in that the function $\I(x)$ has been replaced by another function, called $G(0;xx)$.  This function is expressed in their (3.14) in a form which, after a contour integration, is almost identical to our Eq.~(\ref{di}).  They differ because our $|\g^2(x)-1|$ is replaced by their $|\g^2(x)|$.  Now recall, from the discussion beneath Eq.~(\ref{di}), that the missing $-1$ is necessary to avoid a divergence when integrating over large momenta.  Thus again one sees that our quantity is manifestly finite in the UV, as one expects because of the normal ordering in the defining Hamiltonian.    Indeed the $-1$ in Wick's theorem arose from the plane wave normal ordering.  On the other hand, Ref.~\cite{gjs} states that, in their renormalization scheme, the corresponding divergence must be eliminated using their mass counterterm.

\appendix

\section{Consistency Check: Two-Loop Kink Mass}
For any function $F(x)$, $H\p_F$ is related by a similarity transformation to $H$ and therefore has the same spectrum.  As a result, the energies of all states should be independent of our choice of $F(x)$.  In this appendix we will check that the two-loop ground state kink mass is independent of our choice of $f_2(x)$, and so in particular it will be the same as that found at $f_2(x)=0$ in Ref.~\cite{metwoloop}.

In Ref.~\cite{metwoloop} the kink ground state kink energy $Q=\sum_j Q_j$ was determined using the $\hf$ eigenvalue equation (\ref{ph})
\beq
\sum_{i=0}^j \left(H_{j+2-i}-Q_{\frac{j-i}{2}+1}\right)\vac_i=0 \label{pha}
\eeq
where $\vac_i$ is the order $i$ component of the $\hf$ eigenstate $\vac$.   Similarly the $H\p_F$ eigenvalue equation is
\beq
\sum_{i=0}^j \left(H^F_{j+2-i}-Q^F_{\frac{j-i}{2}+1}\right)\vac^F_i=0 \label{phf}
\eeq
where  $\vac^F_i$ is the order $i$ component of the $H\p_F$ eigenstate $\vac^F$.  If the argument above is correct, then at every order $Q^F=Q$.

The first equation to solve is the $j=0$ equation.  The (\ref{pha}) and (\ref{phf}) equations are clearly identical at $j=0$ as $H_2=H_2^F$.   

\subsection{Leading Order}

At order $j=1$ the original equation was
\beq
0=H_3\vac_0+\left(\frac{\pi_0^2}{2}+\ppin{k}\omega_k B^\dag_k B_k\right)\vac_1
\eeq
and including $f_2$ one instead finds
\beq
0=H^F_3\vac_0+\left(\frac{\pi_0^2}{2}+\ppin{k}\omega_k B^\dag_k B_k\right)\vac^F_1.
\eeq
Subtracting these equations one finds
\beq
g^2\ppin{k}c_{-k}\ok{}^2 \left(B_{k}^\dag+\frac{B_{-k}}{2\omega_k}\right)\vac_0=\left(H^F_3-H_3\right)\vac_0=\left(\frac{\pi_0^2}{2}+\ppin{k}\omega_k B^\dag_k B_k\right)\left(\vac_1-\vac_1^F\right).
\eeq
In Ref.~\cite{metwoloop} it was shown that translation invariance fixes $\vac_0$ up to the kernel of $\pi_0$, by imposing that all states not in the kernel of $\pi_0$ satisfy a recursion relation (\ref{rrs}).  This recursion relation fixes the order $j$ term $\vac_{j}$ in terms of the order $j-1$ term $\vac_{j-1}$.   Including $f_2$ modifies the recursion relation by including the order $j-3$ term, so that $\vac_{j}^F$ is determined in terms of $\vac_{j-1}^F$ and $\vac_{j-3}^F$.  This difference is irrelevant for $j<3$, and so $\vac_1$ and $\vac_1^F$ are equal up to terms in the kernel of $\pi_0$.

This fact simplifies the right hand side while $B_k\vac_0=0$ simplifies the left hand side, leaving
\beq
g^2\ppin{k}c_{-k}\ok{}^2 B_{k}^\dag\vac_0=\ppin{k}\omega_k B^\dag_k B_k\left(\vac_1-\vac_1^F\right).
\eeq
This is easily solved to yield the difference in the eigenvectors of the two Hamiltonians
\beq
\vac_1^F-\vac_1=-g^2\ppin{k}c_{-k}\ok{} B_{k}^\dag\vac_0.\label{vac1}
\eeq

Let us describe the states $\vac$ and $\vac^F$ more systematically.  We expand the states as
\bea
\vac&=&\sum_{i,m,n=0}^\infty \vac_{i}^{mn},\
\vac_i^{mn}=Q_0^{-i/2}\pink{n}\gamma_i^{mn}(k_1\cdots k_n)\phi_0^m\Bd1\cdots\Bd n\vac_0\\
\vac^F&=&\sum_{i,m,n=0}^\infty \vac_{i}^{Fmn},\
\vac_i^{Fmn}=Q_0^{-i/2}\pink{n}\gamma_i^{Fmn}(k_1\cdots k_n)\phi_0^m\Bd1\cdots\Bd n\vac_0.\nonumber
\eea
In other words, all information about a state is contained in the coefficients $\gamma$.   In terms of these coefficients, Eq.~(\ref{vac1}) may be written
\beq
\gamma_1^{F01}(k)-\gamma_1^{01}(k)=-\sqrt{Q_0}g^2 c_{-k}\ok{}.
\eeq

One can check that if $f_2(x)$ is chosen as in (\ref{main}) so as to cancel the leading tadpole, then
\beq
\gamma_1^{F01}(k)-\gamma_1^{01}(k)=\sqrt{Q_0}\frac{V_{\I k}}{2\ok{}}.
\eeq
In Ref.~\cite{metwoloop} it was reported that
\beq
\gamma_1^{01}(k)=\frac{\Delta_{kB}}{2}-\sqrt{Q_0}\frac{V_{\I k}}{2\ok{}}\hsp
\Delta_{ij}=\int dx {\g}_i(x) {\g}\p_j(x)
\eeq
and so we have found that, with this tadpole-canceling choice
\beq
\gamma_1^{F01}(k)=\frac{\Delta_{kB}}{2}.
\eeq
In other words, the eigenvector of the quantum kink Hamiltonian corresponding to the kink ground state is simpler than the corresponding eigenvector of the old kink Hamiltonian.   Physically, we see that the removal of the tadpole eliminates a source of mixing between the one-soliton, zero-meson and the one-soliton, one-meson states.  The remaining mixing is purely kinematic, as it originates in Ref.~\cite{metwoloop} from the translation-invariance of the kink state.

\subsection{Subleading Order: Translation Invariance}

Now let us turn our attention to the next order, $j=2$, corresponding to two loops.   We will again let $f_2$ be an arbitrary function.  In Ref.~\cite{metwoloop} it was shown that the translation invariance of a kink state is equivalent to the recursion relation
\bea
\gamma_{j}^{mn}(k_1\cdots k_n)&=&\left.\Delta_{k_n B}\left(\gamma_{j-1}^{m,n-1}(k_1\cdots k_{n-1})+\frac{\omega_{k_n}}{m}\gamma_{j-1}^{m-2,n-1}(k_1\cdots k_{n-1})\right)
\right. \label{rrs}\\
&&+(n+1)\ppin{k\p}\Delta_{-k\p B}\left(\frac{\gamma_{j-1}^{m,n+1}(k_1\cdots k_n,k\p)}{2\omega_{k\p}}
-\frac{\gamma_{j-1}^{m-2,n+1}(k_1\cdots k_n,k\p)}{2m}\right)\nonumber\\
&&+\frac{\omega_{k_{n-1}}\Delta_{k_{n-1}k_n}}{m}\gamma_{j-1}^{m-1,n-2}(k_1\cdots k_{n-2})\nonumber\\
&&+\frac{n}{2m}\ppin{k\p}\Delta_{k_n,-k\p}\left(1+\frac{\omega_{k_n}}{\omega_{k\p}}\right)\gamma^{m-1,n}_{j-1}(k_1\cdots k_{n-1},k\p)
\nonumber\\
&&\left.-\frac{(n+2)(n+1)}{2m}\int^+\frac{d^2k\p}{(2\pi)^2}\frac{\Delta_{-k\p_1,-k\p_2}}{2\omega_{k\p_2}} \gamma_{j-1}^{m-1,n+2}(k_1\cdots k_{n},k\p_1,k\p_2).
\right.
\nonumber
\eea
As argued above, at $j<3$ the same recursion relation is obeyed by the coefficients $\gamma^F$ of $\vac^F$.

As $\gamma_1^{F01}$ depends on $f_2$, the recursion relation implies that $\gamma_2^{F20}$, $\gamma_2^{F22}$, $\gamma_2^{F11}$ and $\gamma_2^{F13}$ also depend on $f_2$.  Of these, we will see that only $\gamma_2^{F20}$ affects the energy $Q_2^F$.  According to the recursion relation (\ref{rrs}), the contribution to $\gamma_2^{F20}$ from $\gamma_1^{F01}$ is\footnote{Here, the superset symbol indicates that we only write $\gamma_1^{F01}$ contribution.}
\beq
\gamma_2^{F20}\supset 
-\ppin{k}\Delta_{-k\p B}\frac{\gamma_{1}^{F01}(k)}{4}.
\eeq
Therefore we find that $f_2$ shifts $\gamma_2^{20}$ by
\beq
\gamma_2^{F20}-\gamma_2^{20}=-\ppin{k}\Delta_{-k\p B}\frac{\gamma_{1}^{F01}(k)-\gamma_{1}^{01}(k)}{4}=\sqrt{Q_0}g^2 \ppin{k}\Delta_{-k\p B}\frac{c_{-k}\ok{}}{4}.
\eeq

\subsection{Subleading Order: The Eigenvalue Problem}

To calculate the two-loop correction to the ground state energy, $Q_2^F$, we will impose that $\vac$ is an eigenstate of $\hf$ and $\vac^F$ is an eigenstate of $H\p_F$.  The corresponding eigenvalue equations are
\bea
(H_4-Q_{2})|0\rangle_0+H_3|0\rangle_1+(H_2-Q_1)|0\rangle_2&=&0  \label{g2}\\
(H^F_4-Q^F_{2})|0\rangle_0+H^F_3|0\rangle_1^F+(H_2-Q_1)|0\rangle_2^F&=&0  \nonumber
\eea
using the fact that $\vac_0,\ Q_1$\ and $H_2$ are all independent of $f_2$.  Our goal is to show that $Q_2$ is equal to $Q_2^F$.

Subtracting these equations, we find
\bea
D&=&A+B+C\hsp
A=\left(H^F_4-H_4\right)\vac_0\hsp
B=H^F_3\vac_1^F-H_3\vac_1\\
C&=&(H_2-Q_1)\left(\vac_2^F-\vac_2\right)\hsp
D=\left(Q^F_2-Q_{2}\right)\vac_0\nonumber.
\eea
Our goal is to fix $D$.  To do this, we will now evaluate $A$, $B$ and $C$ projected onto $\vac_0$.

Let us begin with $A$.  When $H_4^F$ is normal mode normal ordered, the only term which fails to annihilate $\vac_0$ is the constant term.  
Let us begin by calculating the plane wave normal ordered $H_4^F-H_4$ using the identity
\beq
H^F[\phi(x)]=H[\phi(x)+F(x)]
\eeq
which preserves plane wave normal ordering.  Writing
\beq
H^F_4=\int dx \left(\alpha_4(x) :\phi^4(x):_a + \alpha_2(x) :\phi^2(x):_a+\alpha_0(x)\right)
\eeq
one finds
\bea
\int dx \alpha_0(x)&=&\frac{g^4}{2}\int dxf_2(x)\left(-f^{\prime\prime}_2(x)+\V{2}f_2(x)\right)\\
&=&\frac{g^4}{2}\int dx\ppin{k_1}c_{k_1}{\g}_{k_1}(x)\ppin{k_2} c_{k_2} \ok{2}^2 {\g}_{k_2}(x)
=\frac{g^4}{2}\ppin{k}c_k c_{-k} \ok{}^2
\nonumber
\eea
and
\bea
\int dx \alpha_2(x) :\phi^2(x):_a&=&\frac{g^3}{2}\int dx \V{3} f_2(x) :\phi^2(x):_a\\
&=&\frac{g^3}{2}\int dx \V{3} f_2(x)\left(:\phi^2(x):_b+\I(x)\right)\nonumber\\
&\supset&\frac{g^2}{2}\ppin{k}c_k V_{\I,-k}\nonumber
\eea
where the superset symbol is the restriction of the normal mode normal ordered form to the $c$-number term, which we recall is the only term which does not annihilate $\vac_0$.
The $\phi^4$ term is as in $H_4$, since no $\phi^4$ term appeared at lower order in $g$ and a factor of $f_2$ would necessarily introduce a factor of $g^2$.

Assembling these results, we have found our first contribution.  Projecting $A$ onto $\vac_0$ it is
\beq
A\supset\left(\frac{g^4}{2}\ppin{k}c_k c_{-k} \ok{}^2+\frac{g^2}{2}\ppin{k}c_k V_{\I,-k}\right)\vac_0 \label{aval}
\eeq
where the superset symbol denotes the projection.

Next we turn our attention to $B$.  We may write it as a sum of three terms
\beq
B=(H^F_3-H_3)(\vac_1^F-\vac_1)+
(H^F_3-H_3)\vac_1+
H_3(\vac_1^F-\vac_1).
\eeq
The first term is
\bea
(H^F_3-H_3)(\vac_1^F-\vac_1)&=&g^2\ppin{k_1}c_{-k_1}\ok{1}^2 \left(B_{k_1}^\dag+\frac{B_{-k_1}}{2\omega_k}\right)\left(-g^2\ppin{k_2}c_{-k_2}\ok{2}B^\dag_{k_2}\vac_0\right)\nonumber\\
&\supset&-\frac{g^4}{2}\ppin{k}c_kc_{-k}\ok{}^2\vac_0
\eea
where again the superset is the projection onto the $\vac_0$ direction.  This exactly cancels the first term in (\ref{aval}).  The next term is
\bea
(H^F_3-H_3)\vac_1&\supset&g^2\ppin{k_1}c_{-k_1}\ok{1}^2 \left(B_{k_1}^\dag+\frac{B_{-k_1}}{2\omega_k}\right)
\ppin{k_2}\left(\frac{\Delta_{k_2B}}{2\sqrt{Q_0}}-\frac{V_{\I k_2}}{2\ok{2}}
\right)B^\dag_{k_2}\vac_0\nonumber\\
&\supset&\frac{g^2}{4}\ppin{k}c_k\left(\frac{\ok{}\Delta_{kB}}{\sqrt{Q_0}}-V_{\I,- k}
\right)\vac_0. \label{b2}
\eea
Note that the second term cancels half of the remaining term in (\ref{aval}).
The last term is
\bea
H_3(\vac_1^F-\vac_1)&\supset&\ppin{k_1}\frac{V_{\I k_1}}{2} \left(B_{k_1}^\dag+\frac{B_{-k_1}}{2\omega_k}\right)\left(-g^2\ppin{k_2}c_{-k_2}\ok{2}B^\dag_{k_2}\vac_0\right)\nonumber\\
&\supset&-\frac{g^2}{4}\ppin{k}c_k V_{\I,-k}\vac_0
\eea
which cancels the other half of the remaining term in (\ref{aval}).  Thus we have seen that $B$ cancels all terms in $A$, leaving only the first term in (\ref{b2}).

Finally we calculate $C$
\bea
C&\supset& \left(\frac{\pi_0^2}{2}+\ppin{k_1}\omega_{k_1} B^\dag_{k_1} B_{k_1} \right)
\left(\ppin{k_2}g^2 \Delta_{-k_2\p B}\frac{c_{-k_2}\ok{2}}{4}\frac{\phi_0^2}{\sqrt{Q_0}}\vac_0
\right)\nonumber\\
&\supset&-\frac{g^2}{4\sqrt{Q_0}}\ppin{k} \Delta_{k\p B}c_{k}\ok{}\vac_0.
\eea
This cancels the first term in (\ref{b2}).  We thus conclude that, projecting onto $\vac_0$
\beq
A+B+C\supset 0\vac_0.
\eeq
On the other hand, the projection of $D$ onto $\vac_0$ is $(Q_2^F-Q_2)\vac_0$.  Identifying the two projections, we find
\beq
Q_2^F-Q_2=0.
\eeq
This is an important consistency check of our procedure.  If $H\p_F$ and $\hf$ are similar operators, as we claimed, then although their eigenvectors differ, their eigenvalues must agree order by order.

 \section* {Acknowledgement}

\noindent
JE is supported by the CAS Key Research Program of Frontier Sciences grant QYZDY-SSW-SLH006 and the NSFC MianShang grants 11875296 and 11675223.   JE also thanks the Recruitment Program of High-end Foreign Experts for support.


\begin{thebibliography}{99}
	

\bibitem{dhn2}
  R.~F.~Dashen, B.~Hasslacher and A.~Neveu,
  ``Nonperturbative Methods and Extended Hadron Models in Field Theory 2. Two-Dimensional Models and Extended Hadrons,''
  Phys.\ Rev.\ D {\bf 10} (1974) 4130.
 doi:10.1103/PhysRevD.10.4130

\bibitem{vega}
H.~J.~de Vega,
``Two-Loop Quantum Corrections to the Soliton Mass in Two-Dimensional Scalar Field Theories,''
Nucl. Phys. B \textbf{115} (1976), 411-428
doi:10.1016/0550-3213(76)90497-1

\bibitem{verwaest}
J.~Verwaest,
``Higher Order Correction to the Sine-Gordon Soliton Mass,''
Nucl. Phys. B \textbf{123} (1977), 100-108
doi:10.1016/0550-3213(77)90343-1

\bibitem{menormal}
J.~Evslin and H.~Guo,
``Excited Kinks as Quantum States,''
[arXiv:2104.03612 [hep-th]].

\bibitem{meshape}
J.~Evslin,
``Evidence for the unbinding of the $\phi^{4}$ kink's shape mode,''
JHEP \textbf{09} (2021), 009
doi:10.1007/JHEP09(2021)009
[arXiv:2104.14387 [hep-th]].

\bibitem{mekink}
J.~Evslin,
``Manifestly Finite Derivation of the Quantum Kink Mass,''
JHEP \textbf{11} (2019), 161
doi:10.1007/JHEP11(2019)161
[arXiv:1908.06710 [hep-th]].

\bibitem{mephi4massa}
J.~Evslin,
``The two-loop $\phi^4$ kink mass,''
Phys. Lett. B \textbf{822} (2021), 136628
doi:10.1016/j.physletb.2021.136628
[arXiv:2109.05852 [hep-th]].

\bibitem{tamasstab}
T.~Roma\'nczukiewicz,
``Could the primordial radiation be responsible for vanishing of topological defects?,''
Phys. Lett. B \textbf{773} (2017), 295-299
doi:10.1016/j.physletb.2017.08.045
[arXiv:1706.05192 [hep-th]].

\bibitem{weigelstab}
H.~Weigel,
``Quantum Instabilities of Solitons,''
AIP Conf. Proc. \textbf{2116} (2019) no.1, 170002
doi:10.1063/1.5114153
[arXiv:1907.10942 [hep-th]].


\bibitem{cahill76}
K.~E.~Cahill, A.~Comtet and R.~J.~Glauber,
``Mass Formulas for Static Solitons,''
Phys. Lett. B \textbf{64} (1976), 283-285
doi:10.1016/0370-2693(76)90202-1

\bibitem{memassa}
J.~Evslin,
``Well-defined quantum soliton masses without supersymmetry,''
Phys. Rev. D \textbf{101} (2020) no.6, 065005
doi:10.1103/PhysRevD.101.065005
[arXiv:2002.12523 [hep-th]].

\bibitem{metwoloop}
J.~Evslin and H.~Guo,
``Alternative to collective coordinates,''
Phys. Rev. D \textbf{103} (2021) no.4, L041701
doi:10.1103/PhysRevD.103.L041701
[arXiv:2101.08028 [hep-th]].

\bibitem{mewick}
J.~Evslin,
``Normal ordering normal modes,''
Eur. Phys. J. C \textbf{81} (2021) no.1, 92
doi:10.1140/epjc/s10052-021-08890-7
[arXiv:2007.05741 [hep-th]].

\bibitem{weiszff}
P.~H.~Weisz,
``Exact Quantum Sine-Gordon Soliton Form-Factors,''
Phys. Lett. B \textbf{67} (1977), 179-182
doi:10.1016/0370-2693(77)90097-1

\bibitem{karowskiff}
M.~Karowski and P.~Weisz,
``Exact Form-Factors in (1+1)-Dimensional Field Theoretic Models with Soliton Behavior,''
Nucl. Phys. B \textbf{139} (1978), 455-476
doi:10.1016/0550-3213(78)90362-0

\bibitem{babujianff}
H.~Babujian and M.~Karowski,
``Exact form-factors in integrable quantum field theories: The sine-Gordon model. 2.,''
Nucl. Phys. B \textbf{620} (2002), 407-455
doi:10.1016/S0550-3213(01)00551-X
[arXiv:hep-th/0105178 [hep-th]].

\bibitem{andyff1}
I.~V.~Melnikov, C.~Papageorgakis and A.~B.~Royston,
``Accelerating solitons,''
Phys. Rev. D \textbf{102} (2020) no.12, 125002
doi:10.1103/PhysRevD.102.125002
[arXiv:2007.11028 [hep-th]].

\bibitem{andyff2}
I.~V.~Melnikov, C.~Papageorgakis and A.~B.~Royston,
``Forced Soliton Equation and Semiclassical Soliton Form Factors,''
Phys. Rev. Lett. \textbf{125} (2020) no.23, 231601
doi:10.1103/PhysRevLett.125.231601
[arXiv:2010.10381 [hep-th]].

\bibitem{muri1}
C.~Adam, K.~Oles, T.~Romanczukiewicz and A.~Wereszczynski,
``Spectral Walls in Soliton Collisions,''
Phys. Rev. Lett. \textbf{122} (2019) no.24, 241601
doi:10.1103/PhysRevLett.122.241601
[arXiv:1903.12100 [hep-th]].

\bibitem{muri2}
C.~Adam, K.~Oles, T.~Romanczukiewicz, A.~Wereszczynski and W.~J.~Zakrzewski,
``Spectral walls in multifield kink dynamics,''
JHEP \textbf{08} (2021), 147
doi:10.1007/JHEP08(2021)147
[arXiv:2105.14771 [hep-th]].

\bibitem{gj75}
J.~Goldstone and R.~Jackiw,
``Quantization of Nonlinear Waves,''
Phys. Rev. D \textbf{11} (1975), 1486-1498
doi:10.1103/PhysRevD.11.1486

\bibitem{gjs}
J.~L.~Gervais, A.~Jevicki and B.~Sakita,
``Perturbation Expansion Around Extended Particle States in Quantum Field Theory. 1.,''
Phys. Rev. D \textbf{12} (1975), 1038
doi:10.1103/PhysRevD.12.1038

\end{thebibliography}

\begin{thebibliography}{99}
	
\bibitem{coleman}
S.Coleman,
``Quantum Field Theory Lecture of Sidney Coleman''(2019).World Scientific

\bibitem{me2loopex}
J.~Evslin and H.~Guo,
``Excited Kinks as Quantum States,''
[arXiv:2104.03612 [hep-th]].

\bibitem{salam}
Abdus.Salam
``some Speccuations on the New Resonances,''
Rev. Mod. Phys. \textbf{33}, 426-430 (1961)
doi:10.1103/RevModPhys.33.426

\bibitem{sakurai}
J.~J.~Sakurai,
``Dynamical Approach to the Unitary-Symmetry Mass Formula Based on omega- phi Mixing,''
Phys. Rev. \textbf{132}, 434-441 (1963)
doi:10.1103/PhysRev.132.434

\bibitem{Coleman1964}
S.~Coleman and H.~J.~Schnitzer,
``Departures from the Eightfold Way. 2. Baryon Electromagnetic Masses,''
Phys. Rev. \textbf{136}, B223-B226 (1964)
doi:10.1103/PhysRev.136.B223

\bibitem{Schwinger:1957em}
J.~S.~Schwinger,
``A Theory of the Fundamental Interactions,''
Annals Phys. \textbf{2}, 407-434 (1957)
doi:10.1016/0003-4916(57)90015-5

\bibitem{christlee75}
N.~H.~Christ and T.~D.~Lee,
``Quantum Expansion of Soliton Solutions,''
Phys. Rev. D \textbf{12} (1975), 1606
doi:10.1103/PhysRevD.12.1606

\bibitem{goldstone}
J.~Goldstone and R.~Jackiw,
``Quantization of Nonlinear Waves,''
Phys. Rev. D \textbf{11} (1975), 1486-1498
doi:10.1103/PhysRevD.11.1486










\bibitem{dhn2}
  R.~F.~Dashen, B.~Hasslacher and A.~Neveu,
  ``Nonperturbative Methods and Extended Hadron Models in Field Theory 2. Two-Dimensional Models and Extended Hadrons,''
  Phys.\ Rev.\ D {\bf 10} (1974) 4130.
 doi:10.1103/PhysRevD.10.4130


\bibitem{me2looplett}
J.~Evslin and H.~Guo,
``An Alternative to Collective Coordinates,''
Accepted for publication in PRD,
[arXiv:2101.08028 [hep-th]].



\bibitem{memassa}
J.~Evslin,
``Well-defined quantum soliton masses without supersymmetry,''
Phys. Rev. D \textbf{101} (2020) no.6, 065005
doi:10.1103/PhysRevD.101.065005
[arXiv:2002.12523 [hep-th]].


\bibitem{me2loop}
J.~Evslin and H.~Guo,
``Two-Loop Scalar Kinks,''
[arXiv:2012.04912 [hep-th]].

\bibitem{rajaramanb}
  Rajaraman ~ R.
  ``Solitons and Instantons,''
 Physics Bulletin 34.1(1983):29-29.

\bibitem{rajaraman}
  R.~Rajaraman,
  ``Some Nonperturbative Semiclassical Methods in Quantum Field Theory: A Pedagogical Review,''
  Phys.\ Rept.\  {\bf 21} (1975) 227.
  doi:10.1016/0370-1573(75)90016-2

\bibitem{mekink}
J.~Evslin,
``Manifestly Finite Derivation of the Quantum Kink Mass,''
JHEP \textbf{11} (2019), 161
doi:10.1007/JHEP11(2019)161
[arXiv:1908.06710 [hep-th]].

\bibitem{me1loop}
H.~Guo and J.~Evslin,
``Finite derivation of the one-loop sine-Gordon soliton mass,''
JHEP \textbf{02}, 140 (2020)
doi:10.1007/JHEP02(2020)140
[arXiv:1912.08507 [hep-th]].


\bibitem{rebhan}
  A.~Rebhan and P.~van Nieuwenhuizen,
  ``No saturation of the quantum Bogomolnyi bound by two-dimensional supersymmetric solitons,''
  Nucl.\ Phys.\ B {\bf 508} (1997) 449
  doi:10.1016/S0550-3213(97)00625-1, 10.1016/S0550-3213(97)80021-1
 [hep-th/9707163].


\bibitem{local}
A.~S.~Goldhaber, A.~Litvintsev and P.~van Nieuwenhuizen,
``Local Casimir energy for solitons,''
Phys. Rev. D \textbf{67} (2003), 105021
doi:10.1103/PhysRevD.67.105021
[arXiv:hep-th/0109110 [hep-th]].



\bibitem{cahill76}
K.~E.~Cahill, A.~Comtet and R.~J.~Glauber,
``Mass Formulas for Static Solitons,''
Phys. Lett. B \textbf{64} (1976), 283-285
doi:10.1016/0370-2693(76)90202-1


\bibitem{mewick}
J.~Evslin,
``Normal Ordering Normal Modes,''
Eur. Phys. J. C \textbf{81} (2021) no.1, 92
doi:10.1140/epjc/s10052-021-08890-7
[arXiv:2007.05741 [hep-th]].

\bibitem{me2stato}
J.~Evslin,
``Constructing Quantum Soliton States Despite Zero Modes,''
[arXiv:2006.02354 [hep-th]].







\bibitem{gjs}
J.~L.~Gervais, A.~Jevicki and B.~Sakita,
``Perturbation Expansion Around Extended Particle States in Quantum Field Theory. 1.,''
Phys. Rev. D \textbf{12} (1975), 1038
doi:10.1103/PhysRevD.12.1038

\bibitem{jackiw77}
R.~Jackiw,
``Quantum Meaning of Classical Field Theory,''
Rev. Mod. Phys. \textbf{49} (1977), 681-706
doi:10.1103/RevModPhys.49.681



\end{thebibliography}
 \end{document}